# Twigraph: Discovering and Visualizing Influential Words between Twitter Profiles

Dhanasekar S[1] and Sudharshan Srinivasan[1]

[1] Chennai, India
dhanasekar312213@gmail.com

**Abstract.** The social media craze is on an ever increasing spree, and people are connected with each other like never before, but these vast connections are visually unexplored. We propose a methodology *Twigraph* to explore the connections between persons using their Twitter profiles. First, we propose a hybrid approach of recommending social media profiles, articles, and advertisements to a user. The profiles are recommended based on the similarity score between the user profile, and profile under evaluation. The similarity between a set of profiles is investigated by finding the top influential words thus causing a high similarity through an Influence Term Metric for each word. Then, we group profiles of various domains such as politics, sports, and entertainment based on the similarity score through a novel clustering algorithm. The connectivity between profiles is envisaged using word graphs that help in finding the words that connect a set of profiles and the profiles that are connected to a word. Finally, we analyze the top influential words over a set of profiles through clustering by finding the similarity of that profiles enabling to break down a Twitter profile with a lot of followers to fine level word connections using word graphs. The proposed method was implemented on datasets comprising 1.1 M Tweets obtained from Twitter. Experimental results show that the resultant influential words were highly representative of the relationship between two profiles or a set of profiles.

## 1 Introduction

The important characteristic of a successful social media is its large, engaged user base. Hence, every social media tries to improve its user base. Twitter is one such popular social media site providing microblogging service that has been an important representative of people's personal opinion in the past decade [1]. People use Twitter to share and seek information ranging from gossips to the news [26,27], as its range of connectivity far greater than any other medium. Now Twitter has around 317 million users worldwide and about 500 million tweets posted per day. Though it has tons of information with monumentally large user-base, it is practically impossible for a user to find fellow users who share a common interest manually. There is a need for an efficient user suggestion system that can group users with similar interests. An automated suggestion system [3] helps a user to find other users with similar interests, thus acquiring and sharing knowledge about a particular domain.

Now after an efficient recommendation system is built, a user develops his follower list. This list is built gradually or radically depending on the user's status, and popularity resulting in the accumulation of the followers. These followers would have followed the user based on his nature of the user’s tweets. By nature, here we mean the topics used in the tweets. If the user is a guitarist and tweets were highly concentrated on acoustics, electrics and the brands of guitar, the followers of that user would probably have these topics in the majority. But when the user is a worldwide popular celebrity or politician, the nature of tweets may span several topics ranging from philosophy to cinema. Hence the followers of such a user may have followed that user for a range of topics found in his tweets. Though this is obvious, what if there is a way to find the important or influential words between a user and his follower group causing a person to be a follower. This method called *Twigraph* would enable to visualize the connectivity across profiles through words and vice versa (connectivity across words through profiles).

To summarize,

- We take approximately 3000 tweets of various users of domains like sports, politics, philosophy and education from Twitter. We also take a large number of news and advertisement articles available online. Subsequently, we analyze, pre-process and store them efficiently.
- A profile under evaluation (user profile) is chosen, and top profiles similar to that of the user profile based on his nature of tweets are found (Explained in the upcoming sections).Article and advertisement suggestions are also made.
- Then, we analyze the top influential words between a profile and the gradually evolving user group (user profile and his followers) using Influence Term Metric (ITM) and a variant of clustering algorithm (proposed in Section 6).

The paper is organized as follows. Section 2 deals with the related works about the usage of Twitter as a social media data set in performing various tasks like document clustering and topic modeling. Section 3 talks about the collection of data from Twitter and preprocessing it. Section 4 gives a glimpse about the profile (Twitter profile) modeling and the distance measures used for finding the distance between profiles with an example. Section 5 explains the proposed hybrid suggestion system for advertisements, articles, and users. Section 6 performs a new clustering technique based on the user profile(query). Section 7 explains the method of finding influential words between a set of profiles using Influence Term Metric with great details and finally Section 8 illustrates the visualization techniques namely *word graphs* to envisage the connection between profiles in words and we finally conclude with future works in Section 9.

## 2 Related Works

Analysis and recommendations for Twitter have been a widely researched topic. Implementation of techniques ranging from simple text mining to more complex learning algorithms has been proposed. The various ways and fields where Twitter data is used are summarized in the following.

**Twitter as a source for text mining.** Twitter is seen as an instant and short form of communication for users to share and seek information.Twitter provides large potentially useful data for purposes such as sentiment analysis, opinion mining, recommender systems, etc. The usage of social media like Twitter to share and seek day-to-day information and how this information can be analyzed is done in [1]. The increased usage of microblogging service in the recent years and how a large amount of data present in the form of tweets can be effectively used for text mining is well described in [2].

**Twitter recommendations.**

**Users**. Twitter has a monumentally large number of users. It is practically impossible to find users with similar interests manually as discussed in the introduction and hence there have been many works in building a user suggestion system. One such paper which describes how users with similar intentions connect with each other is shown [3]. Though Twitter provides a lot of information, the problem of finding followers with similar interests using various recommendation techniques is compared and contrasted in [4].

**News and Advertisements.** Twitter profile data can be extracted, analyzed and used for not only finding users with similar interests but can also be extended to suggest most similar news articles and advertisements. Recommending a dozen of articles from millions saves the user a lot of time. One such recommender system using click behavior and web history to predict news articles is done [5]. A novel recommender system using real-time Twitter data to recommend news articles is proposed [6]. Leveraging Twitter feed by user modeling and analyzing temporal dynamics of profiles for recommending news articles is conceptualized [7] while using other information like the location in addition to user modeling to increase the efficiency of news recommendation is also performed [8,9]. While the above works were concerned with the users, the recommendation technique also enables companies and brands to enforce personalized marketing to their potential customers [10].

**Twitter for Forecasting.** The Twitter data in the form of tweets can not only be used for finding users and articles with similar interests, but it can also be used efficiently for forecasting. Twitter data based sentiment analysis can predict the mood of the users in the social media and thus enable a key factor for prediction. Two such works [11,12] proposes the use of sentiment analysis on Twitter corpus data to effectively forecast elections results in advance.

**Twitter profile modeling and similarity.** A Document is modeled by identifying the keywords in it. Similarly, it can also be applied to a Twitter profile to find the words that are representative of the profile. TF-IDF have been used for finding word relevance and feature selection of terms in a document [13,14] where the keywords of a document have been identified [13]. After modeling, it is worthwhile to find the similarity between two documents. This similarity can be achieved through a variety of similarity metric measures. Analysis of various distance measures for finding distances between two documents and the advantages and disadvantages of the same are emphasized [15].

**Document clustering.** After finding the keywords of a document, finding distances across documents, the possible next idea is to group similar documents together, which is achieved through clustering. An analysis of the various clustering methods for documents is done in [16,17]. To be specific, it provides with a comparative study of agglomerative hierarchical clustering and K-means clustering. Clustering enables the searching of documents efficiently, and a technique for clustering text documents for browsing large document collections is done [18]. The TF-IDF and the clustering approach together for clustering English text documents that are more relevant are performed in [19]. The clustering of documents using TF-IDF scores at word levels for classifying the sentiment of the document as positive or negative is done [20]. In another work, the TF-IDF scores of words of different documents to perform clustering for the application of finding relevant search results for the user query using cosine distance is explained [21]. The above works details about document modeling, similarity, and clustering. An interesting work that uses the combination of TF-IDF and centroid-based clustering for summarizing multiple documents is conceptualized [22], and finally, the semantic similarity between texts using IDF as a metric is done in [23].

To summarize, there have been many works using the combination of TF-IDF with clustering for document topic modeling, stop words removal and document clustering for topic classification. We focus on finding the similarity between documents (profiles) and go to the next level in finding the words that are impactful between the two documents or a document with a set of another document causing the similarity. We propose a term Influence Term Metric (ITM) based on TF-IDF and a variant of clustering algorithm to achieve our case and finally propose a visualization paradigm in the form of word graphs and word paths that envisage the connection between Twitter profiles in the form of words and the profiles that are connected to a word.

## 3 Data Collection and Preprocessing

In this section, we first describe the Twitter data and then the process through which we collected tweets related to the domains of politics, sports, entertainment, education, and philosophy (Section 3.1).

### 3.1 Twitter Data Collection

Figure 1 illustrates the stage-wise filtering in the extraction of Twitter profiles data (pipeline). The data is obtained by using the official Twitter API, TweePy [24]. First, a huge list of Twitter profiles is created. Each profile has a large number of tweets. Second, a Language filter is applied to extract only the profiles that share the common language. Language filter ensures consistency across Twitter profiles in the choice of words as the same word written in the different language is perceived different. Then a Domain filter is applied to extract profiles that come under five chief domains such as politics, sports, philosophy, entertainment and education. Finally, Popularity filter is applied to ensure that the profiles have a minimum number of tweets and have a considerable number of followers. We extract approximately 3000 Tweets per user for various Twitter profiles of various domains. At an average of 120 characters per tweet, our dataset comprises 1.1M tweets with 130M characters approximately.

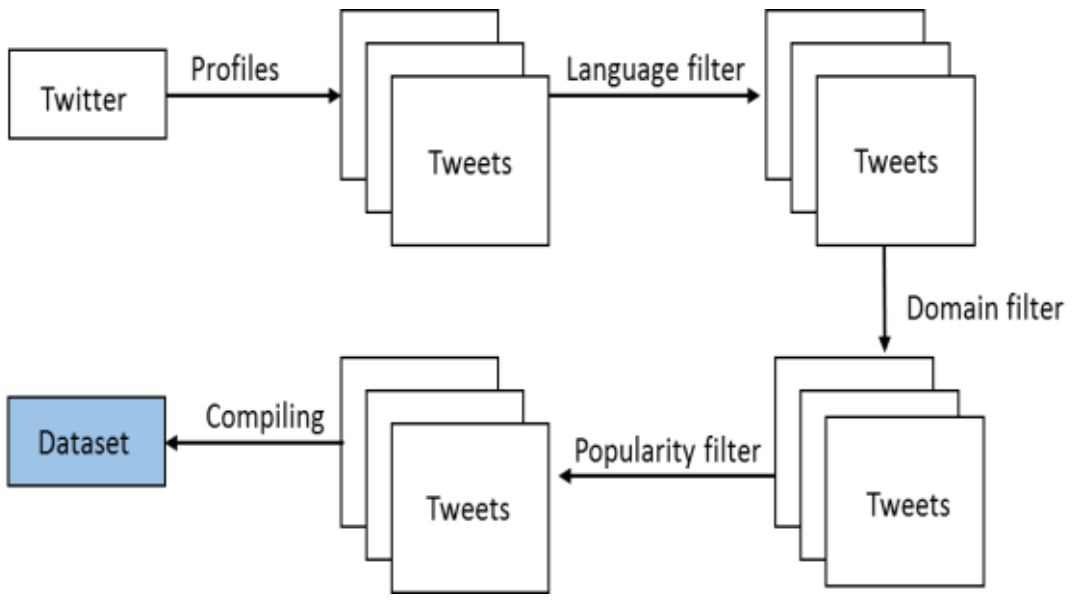

**Fig. 1.** Stage-wise extraction of Twitter data.

### 3.2 Preprocessing

To transform the data into an efficient format and to get rid of unwanted words in a tweet that isn't required for our methodology, we preprocess the data. Here, Preprocessing effectively deals with replacing or removing a word. To avoid the case where the same word occurs twice in different forms, but convey the same because of acronyms, we preprocess the data with an acronym dictionary, by converting acronyms to their full form. For example, lol is translated to "laughing out loud". While in removing, words containing symbols such as '@,' #' and words that are just numbers are removed. The 1.1M such preprocessed tweets are carefully documented based on their domains. For our proposed methodology, it is essential that the tweets aren't just arranged as stand-alone strings but rather as profiles. Hence each tweet from a particular profile is merged to form one single profile document, which makes up a comprehensive dataset.

## 4 Profile Modeling and Similarity

In this section, we first discuss profile modeling by finding the most representative words of that particular profile (Section 4.1) and then we find the similarity between profiles (Twitter profile) using the suitable distance metric (Section 4.2) by demonstrating a comparison between sets of profiles.

### 4.1 Profile Modeling

Profile modeling refers to the top words of a profile, by top we mean the most important words that often reflect the profile (document). If this importantness is quantified using the count of the words, then the articles and conjunctions like 'and' and 'if' becomes the top words most of the times. But these are mere stop words that don't reflect the characteristic of the profile in any way. Hence we calculate the TF-IDF of each word in the corpus. TF-IDF is the feature selection approach used in case of documents to find the words that are highly representative of the document and ignoring stopwords that don't convey any meaning [14]. It is also used to find what words in a document that is favorable to use in a search query to fetch that document [13]. TF-IDF can be successfully used for stop words filtering in various subject fields including text summarization and classification.The document modeling for sample documents is shown in Appendix A.

## 4.2 Profile Similarity

Now that the top words in a profile are identified using TF-IDF scores, the similarity between two profiles (Twitter profiles) can be calculated. There are various distance metrics such as Euclidean, Manhattan, Cosine, etc. to calculate this similarity. But Cosine distance is the apt choice [15] here as it avoids the bias caused by different document lengths evident from the TF-IDF scores as the inner product of the two vectors (sum of the pairwise multiplied elements) is divided by the product of their vector lengths. The resultant score is a value between 0 and 1. The similarity score is obtained by subtracting this value from 1.

$$\textit{Cosine distance} = (d1d2)/||d1||||d2|| \tag{1}$$

Where d1 and d2 are the vectors of the same length.

$$\textit{Cosine similarity} = 1 - (d1d2)/||d1||||d2|| \tag{2}$$

While calculating the distance between the user profile (query) and an another profile, the words contained in the user are alone taken. The TF-IDF scores corresponding to the words present in the user profile is used for cosine distance calculation. That is, the words present in the user profile is searched in the comparing profiles. Next, the TF-IDF scores of these words in comparing profiles is compared with the TF-IDF scores of the user profile.

**Table 1.** TF-IDF of user profile terms (HillaryClinton)

| Terms | TF-IDF |
|---|---|
| immigrant | >0 |
| election | >0 |
| federal | >0 |
| twitter | >0 |
| washington | >0 |

**Table 2.** TF-IDF of profile 1 terms(realDonaldTrump)

| Terms | TF-IDF |
|---|---|
| immigrant | >0 |
| election | >0 |
| Federal | >0 |
| twitter | >0 |
| washington | >0 |

**Table 3.** TF-IDF of profile 2 terms(katyperry)

| Terms | TF-IDF |
|---|---|
| immigrant | 0 |
| election | 0 |
| federal | 0 |
| twitter | >0 |
| washington | >0 |

Table 1 shows us the TF-IDF of the sample words (randomly chosen) in the user profile document. Table 2 and 3 shows us the sign of TF-IDF of the sample words (chosen from HillaryClinton's profile) in the profile 1 (realDonaldTrump). and profile 2 (katyperry) documents. A glimpse of handful picked terms from the user profile (HillaryClinton) is compared in both profiles 1 and 2 (realDonaldTrump and katyperry). In Table 2, all the words that are chosen from the user profile (HillaryClinton) have a positive TF-IDF scores, which indicate the mere presence of that word in profile 1 (realDonaldTrump), but not in Table 3 with only two words having a positive score. Though the number of words chosen here is a fraction of the total words of that profile, it is found that profile 1 (realDonaldTrump) is more similar to the user profile(HillaryClinton) than that of profile 2(katyperry) using cosine distance. Like this, if the process is repeated for all the profiles with the user profile(query), the similarity can be sorted in decreasing order to find the most similar profile to the user profile and second most similar and so on.

## 5 Hybrid Suggestion System

A hybrid Suggestion system is a one which is capable of suggesting articles, advertisements, and profiles for a given user profile. Users would get their most similar profiles based on the nature of their profile to follow and most similar articles like news, entertainment, and research to read. Companies, on the other hand, would want to identify their potential customers by analyzing social media demographics instantly. We describe suggesting users, articles and users for companies in the upcoming subsections.

### 5.1 Suggesting Users for user

Suggesting users for a given user profile is explained in the previous section in great detail. As aforesaid, finding similar users can be a cumbersome task if done manually and automating this would be of utmost importance to gather people of similar interests. One of the works detail the architectural overview, and the graph recommendation algorithms for finding Twitter followers [3]. Another work suggests that, though Twitter provides a lot of information, one of the drawbacks is the lack of an effective method to find fellow users to follow and make friends [4]. As detailed in the previous section, we find the top 3 users for a profile. Table 4 gives the top 3 users for each of the profiles namely HillaryClinton, and rihanna. The top 3 users for HillaryClinton are THEHermanCain, realDonaldTrump, and GovMikeHuckabee who happen to share the common domain namely politics. For rihanna, the top 3 users are lenadunham, ddlovato and souljaboy who are singers in entertainment industry, since rihanna is a singer.

**Table 4.** Similar Users for each user profile

| User profile | Top 3 users | Rank |
|---|---|---|
| HillaryClinton | THEHermanCain | 1 |
| | realDonaldTrump | 2 |
| | GovMikeHuckabee | 3 |
| rihanna | lenadunham | 1 |
| | ddlovato | 2 |
| | souljaboy | 3 |

### 5.2 Suggesting Articles

**Table 5.** Similar articles for each user profile

| User profile | Top 2 articles | Rank |
|---|---|---|
| HillaryClinton | Alex Wallace to head Washington | 1 |
| | Brazil spied on US diplomats | 2 |
| Rihanna | Actors to watch this fall | 1 |
| | 10 best dresses in movie history | 2 |

Social media users would find tons of articles in their daily life. These articles may range from cinema, education, research and so on. Most of these articles would not be of much interest to a particular user as his range of interests may be limited to a small domain. So the suggestion system proposed for users can also be extended to suggesting articles to a particular user to suit his needs through deliberate preprocessing. One of the prediction system uses click behavior and web history to predict articles [5]. Another news suggestion system using real-time Twitter data as and when tweeted is also done [6]. Apart from using the Twitter data alone, other information like the geographic location to increase the news recommendation efficiency is also proposed [8-9]. Table 5 gives the top 2 articles for each of the profiles namely HillaryClinton and rihanna.

### 5.3 Suggesting Users for Companies

For companies or brands, personalized ad targeting based on the interest shown by the users would prove efficient as there is more possibility of a relevant user turning into a potential customer than a common user. Companies can use Twitter to display the most relevant ads to the respective users based on the nature of their tweets [10], performing personalized ad targeting. Table 6 gives the top 2 users for each of the brands namely Nike and BBC. The top 2 users for Nike happens to be footballers. While for BBC, a British news channel, the top user is number10gov which is the handle of UK prime minister; the second top user is a British referee.

**Table 6.** Similar users for each company/brand

| Brand | Top 2 users | Rank |
|---|---|---|
| Nike | Alex Morgan | 1 |
| | Wayne Rooney | 2 |
| BBC | number10gov | 1 |
| | Graham Scott | 2 |

## 6 User Profile based Single Source Clustering

After extracting tweets from different profiles, computing importance of terms in each profile, calculating similarity with each other (user profile and other profiles) and ranking them accordingly to each profile, we arrive at the final and key step, grouping similar profiles with each other. A comparative analysis of K-means and hierarchical clustering for documents is done [16]. An analysis of the various document clustering methods by showing the feature selection methods, similarity measures and evaluation measures of document clustering is done [17]. The use of clustering documents for browsing large document collections is presented in [18], document clustering for fetching relevant English documents in [19]. Clustering is also used for sentiment analysis in predicting the mood as positive or negative [20]. Finally, clustering is used for extracting key sentences from a paragraph based on the user query using the combination of TF-IDF and Cosine distance [21].

A key variant of the Hierarchical clustering algorithm is proposed for the grouping of profiles such that, the grouping does not take place across different profiles, but always showing prominence only on the user profile. Hence, a single source clustering algorithm is proposed as to focus on the user profile.

**Single Source Clustering Algorithm**

```
C            ←An array that stores TF-IDF scores of incoming profiles into the cluster
Profiles     ←TF-IDF scores of all profiles for each word in user profile
Profiles(q)  ←TF-IDF scores of words in the user profile
Profiles(i)  ←TF-IDF scores of user profile words in  profile i
```

```
N, i=500
Start:
      Dist,P = inf
      Count=0
Loop:
      d=cosine distance(Profiles(q),Profiles(i))
      if (d < dist) then
            dist=d
            P=i
      Profiles(i) ← Profiles(i+1)
      count ← count +1
      If (count < N) then
            goto Loop
profiles(q) ← (Profiles(q) + Profiles(P))/2
C.append(P)
Profiles.remove(P)
N ← N - 1
If (N  > 0) then
      goto Start
```

As mentioned in the algorithm, *profiles*(q) is the user profile (profile under query) that contains the TF-IDF scores of terms in that profile while *profiles* is the list with all the profiles that contain TF-IDF scores of user profile terms in each profile. All the remaining profiles in profiles except *profiles*(q) is iteratively compared with profiles(q) for shortest distance. Eventually, the closest profile to the user profile (which is also seen in Table 4 of the previous section) is added to the cluster. Then the TF-IDF scores of these profiles are averaged (Centroid). The centroid calculation enables the score of terms that occurs in both the profiles to be rewarded and scores of terms not in incoming profile (closest) to be penalized. The algorithm continues by finding the closest profile to the gradually forming cluster until *profiles* list is exhausted. In the first iteration, query user profile (Hilary) is merged with the most similar profile (THEHermainCain), thereby forming the first cluster. This cluster is formed, as the TF- IDF scores of these two profiles were similar enough for the profile (THEHErmainCain) to be ranked first to user profile. Now the centroid of these two profiles is calculated as the cluster center. The next closest profile to this centroid score is then added to the cluster. By closest, we mean the next profile (realDonaldTrump) with the TF-IDF scores of those words in the user profile closest to the centroid than any other profile. This way, the cluster is aggregated.

From Table 7, it is clear that the order in which the profiles enter into the cluster is not the same as the closest neighbors given in the previous section. It is because of the gradual change in scores of terms by averaging out the scores. The first profile to enter the cluster is the user profile's closest neighbor, the second profile to enter is not the second closest neighbor, but the closest to the both the profiles in the cluster combined. This way, the diversity of terms is increased. For example, if the user profile is a politician who tweets only about politics, the first profile to enter will obviously be a politician. However, if the profile that enters has some percent of tweets related to the entertainment industry, the profiles related to entertainment industry soon has a chance to enter the cluster. It is evident from the word cloud generated based on the top influential words between a user group cluster and profile.

**Table 7.** Order of Entry for clustering(Hillary Clinton)

| Profile | Entry number |
|---|---|
| THEHermanCain | 1 |
| realDonaldTrump | 2 |
| GovMikeHuckabee | 3 |
| newtgingrich | 4 |
| PeterBale | 5 |

Figures 2, 3, and 4 illustrate the word clouds [25] formed based on the top impactful words between the closest incoming profile and the current formed cluster using the algorithm mentioned. The word clouds are made of words based on their importance between the current cluster and the incoming closest profile, calculated using the ITM (described in the next section). Figure 2 gives the word cloud between user profile (HilaryClinton) and the closest profile to the user profile (THEHermaineCain). Almost all of the words denote about politics as these two profiles are politicians. Figure 3 illustrates the top words between the existing cluster (around 302 profiles including HilaryClinton and THEHermaineCain) and SpeakerRyan, while Figure 4 illustrates the influential words between JimmyFallon and the existing cluster. It can be noted that the

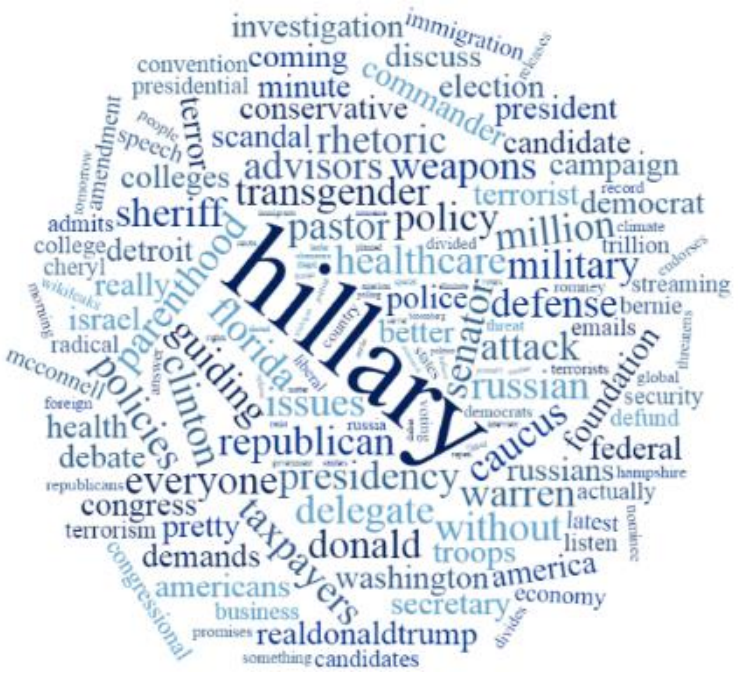

**Fig. 2.** Top Influential words between HilaryClinton and THEHermaineCain

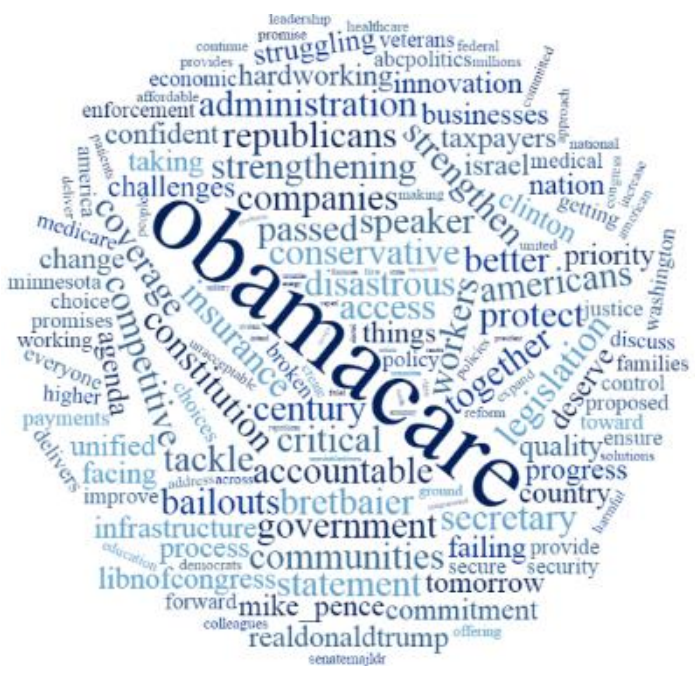

**Fig. 3.** Top Influential words between the current Cluster and SpeakerRyan

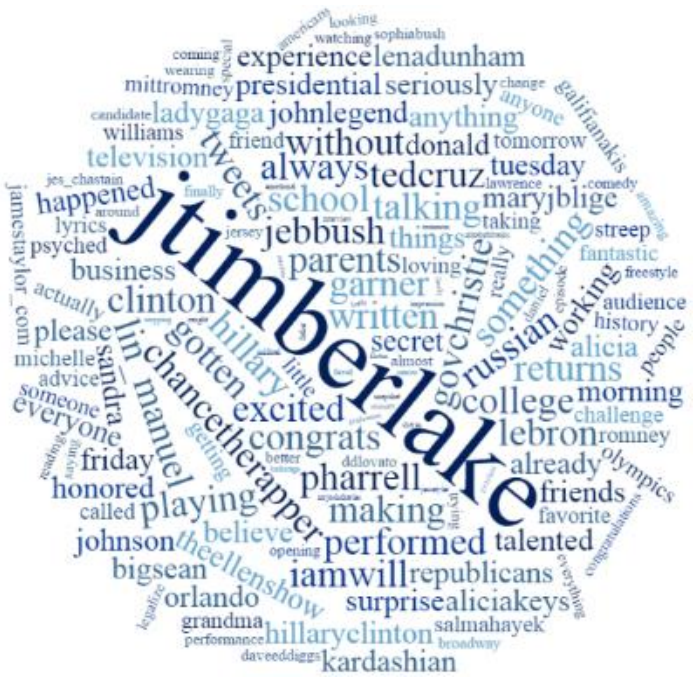

**Fig. 4.** Top Influential words between the Cluster and JimmyFallon

first two Figures are almost about politics while the last Figure is about entertainment.

## 7 Finding the Top Influential words between Profiles

In this section, we analyze the formation of clusters by finding influential words using *Influence Term Metric (ITM)*.

**Influence Term Metric**. The *Influence Term Metric* for a term uses the TF-IDF scores of term in individual profiles and the global IDF score of the term. While the TF-IDF score of a term in a profile indicates the relative importance of that term in the concerned profile, the IDF score of that term indicates its importance in the entire corpus. The *Influence Term Metric* of a term indicates the importance of that term between a set of profiles. Here by 'set' we mean two profiles or between a cluster and profile.

$$ITM(X_{MN}) = T_{XM} * I_X * T_{XN} \tag{3}$$

$ITM(X_{MN})$ -> Influential Term Metric of term 'X' across profiles M and N
$T_{XM}$ -> TF-IDF of term 'X' in document M

$T_{XN}$ -> TF-IDF of term 'X' in document N

$I_X$ -> IDF of term 'X' in the corpus.

The efficiency of the ITM for various cases of $T_{XM}$ and $T_{XN}$ are as follows.

**Case 1**: If both $T_{XM}$ and $T_{XN}$ is low, then obviously $I_X$ is low as the term is less important in the corpus, then the ITM of that term is very low and proves to be less influential between the two documents.

**Case 2**: If $T_{XM}$ is low and $T_{XN}$ is high, that means that term has a high TF in document N overshadowing it's relatively less $I_{X,}$ and in this case the ITM is mediocre. The same case occurs for the contrary case of high $T_{XM}$ and low $T_{XN}$.

**Case 3**: If both $T_{XM}$ and $T_{XN}$ is high, indicating that the term is of high importance in both the documents and $I_X$ is high, then the ITM is high. In a rare case, if $T_{XM}$ and $T_{XN}$ are high, but $I_X$ is low (for the case where TF of that term overshadows the $T_{XM}$ and $T_{XN}$), the ITM becomes mediocre as the $I_X$ is low.

**Table 8.** Top 3 influential words between cluster and incoming profile(Distance based)

| EC | IP | Top 3 words | RC | ITR |
|---|---|---|---|---|
| HillaryClinton | THEHermanCain | hillary<br>bernie<br>obamacare | I | 1 |
| I | realDonaldTrump | america<br>mike_pence<br>pennsylvania | II | 2 |
| II | GovMikeHuckabee | israel<br>abolish<br>medicare | III | 3 |
| LXXIV | Stephen_Curry | science<br>newyorker<br>universities | LXXV | 75 |
| CIII | Reillymj | climate<br>warming<br>global | CIV | 104 |
| CXXII | faisalislam | election<br>government<br>amendment | CXXIII | 123 |
| CXLIII | Number10gov | secretary<br>investment<br>economy | CXLIV | 144 |
| CCLXVII | DjokerNole | tennis<br>practice<br>tournament | CCLXVIII | 268 |

**Notations used**: EC- Existing Cluster, IP- Incoming Profile, RC- Resultant Cluster, ITR- Iteration.

Table 8 shows the entry of profiles into the cluster based on the distance. That is the profiles with the shortest distance to the cluster formed so far enters the cluster. Our focus (as mentioned in the abstract) is to find the top influential words that caused the incoming profile to be the one with the shortest distance to the cluster. In other words, these are the common words between the cluster and the incoming profile and also were representative of the profile or cluster they belong. These words are found using the ITM. The different cases of TF-IDF scores of the words in the cluster and the incoming profile and their impact on the ITM are already detailed. Let us discuss them with few examples. In the first iteration, the term "flotus" has a high TF-IDF score in the user profile (HillaryClinton) and a good IDF score. It could have made it to the Top 5 words list had it had a good TF-IDF score in closest incoming profile resulting in the formation of the cluster (THEHermanCain). But the TF-IDF score of the term "flotus" in the incoming profile is 0, making the ITM of that term 0. The term "hillary" is the one with the highest TF-IDF score in the user profile (HillaryClinton), it also has a higher score in the closest profile (THEHermanCain) and the IDF of that term is high enough for it to be the top most influential word (Highest ITM). On the other hand, the term "obamacare" has a higher TF-IDF in incoming profile (THEHermanCain) than the term "bernie" and the IDF of "obamacare" is higher than "bernie" too. But it's TF-IDF in the user profile (HillaryClinton) is too low to beat the ITM score of "bernie" making it the second most influential between them. Many such scenarios can be explained, and the result is the words that have global importance and also importance in both the individual profiles.

On the other hand, the similar function can be performed for displaying the top influential words between the incoming profile and the existing cluster but not based on the distance but based on chronological order of entry (building of followers) to decompose a Twitter account using *Twigraph*. This is explained in Appendix B.

This finding of influential words between profiles helps in grouping a large user base in social media together at the finest level, that is in words they have used in the social media. It also helps to analyze the gradual change in the topic or choice of words a profile has and the impact it has in connection with the other profiles. The relationship between two Twitter profiles in words can be visualized using *word graph* and *word path* combinely forming *Twigraph*. *Word graph* denote the connection between two profiles in words, which in turn can be used to find profiles that are connected to a word. *Word path* denotes the tracing the *word graph* from one profile to another through to obtain a series of words. The word graph is explained in detail in the next section.

## 8 Visualizing Word Graphs

In this section, we provide a visual representation and analysis of our proposed methodology *Twigraph*.

**Notations used in Figure 5 and 6:**

1. **Squared letter:** Indicates an individual profile
2. **Squared number:** Indicates the cluster formed at that particular iteration.
3. **Oval:** Represents the word connecting two profiles or a profile and cluster.
4. **Blow-up bubble:** Represents the components of that particular cluster, i.e., the cluster in its previous iteration
5. **Blue line**: Indicates that the word connecting two profiles or a profile and cluster, features among the top 20 words shared between them.
6. **Red line:** Indicates that the word connecting the two profiles or a profile and cluster, does not feature among the top 20 words shared between them.

7. **Dashed line**: Represents the word that connects an incoming profile to a component in the blow-up bubble( used to identify if the previously entered profile share that word with the newly incoming profile.

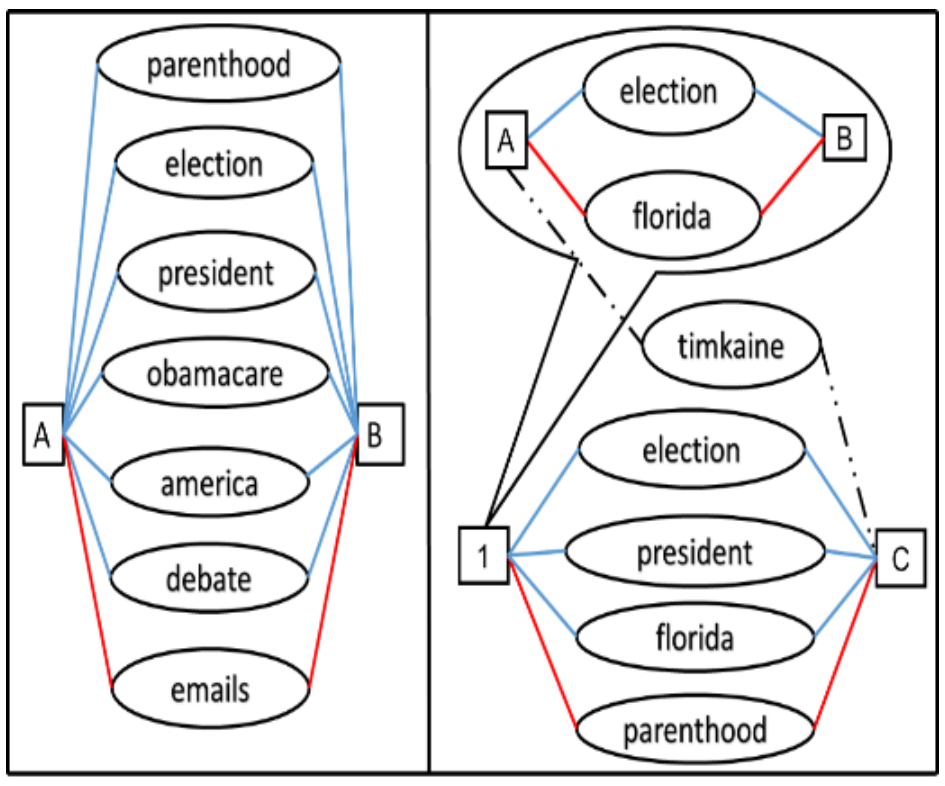

**Fig. 5.** Word graph for Iteration 1 and 2 of Clustering (Table 8)

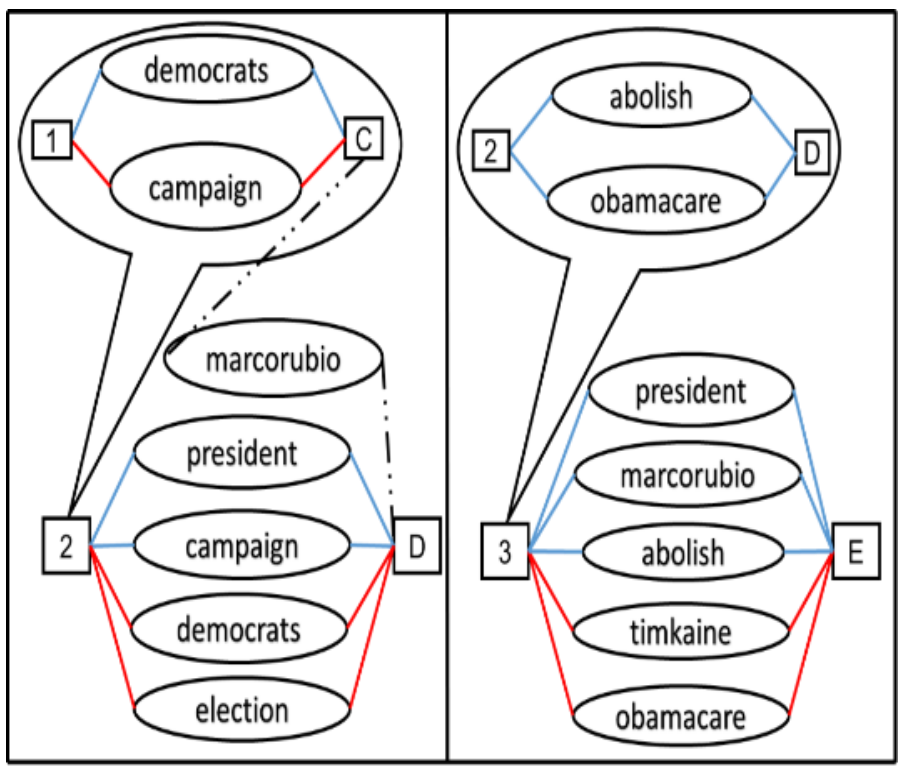

**Fig. 6.** Word graph for Iteration 3 and 4 of Clustering

A lot of interesting analysis could be made out of Table 8 and Fig 6 and 7 (word graph). In general, there are four possible scenarios that could be observed on the word path during the process of clustering:

**Decreasing significance.** Taking the word "parenthood " as an example, we observe that it belongs in the top 20 influential words shared between the user profile (HillaryClinton) and her closest profile (THEHermanCain) as there is a high TF-IDF of that word in both their tweets. But when other profiles start coming into the equation as we progress with the clustering process, the word loses it's significance and moves out of the top 20 list because of its low usage amongst the newly clustered profiles. The same could be said for the word "democrats" . HillaryClinton and THEHermanCain use that word with a very high frequency, and hence it makes the top 20 list of words shared between them. But due to its relatively low usage amongst the next incoming profiles, the word loses its significance.

**Increasing significance.** What happens if the influence for a particular word is very low for the first few incoming profiles but increases over several iterations? This leads to our second scenario where there is an increase in significance for a particular word with the progression of the clustering process. The word "marcorubio" can be used to describe this scenario perfectly. There is no usage of that word from THEHermanCain and hence it doesn't make the list of influential words. But there is some level of usage from realdonaldtrump and GovMikeHuckabee who are the third and fourth profile respectively. This ensures that the word enters the list of common words between existing cluster and incoming profile but not enough to push it to the top 20 influential words list. The word finally makes the top 20 list with the entry of newtgingrich as he had heavily used it in his tweets. The word "medicare" is another similar example.

**Maintaining significance.** This scenario is commonly observed when the usage of a word remains reasonably constant across several incoming profiles. One such example is the word "president". Since the first four closest profiles to HillaryClinton are all politicians, the word "president" is a common occurrence among their tweets. Hence it consistently features in the top 20 list across the first four iterations. Similarly, the word "america" follows the same scenario of maintaining its significance across iterations.

**Oscillating Significance.** This is the final scenario which can be observed from the progression of the clustering process. As the name suggests, it occurs when a particular word oscillates between high and low influential across several iterations. Every word will eventually follow this pattern if we are to increase the range of our observations across many iterations, but we are more interested with oscillations within a short range of iterations. For instance, the word "timkaine" helps us better understand this scenario. THEHermanCain didn't use this words in his tweets, resulting in it's absence from the list of common words for iteration one. But realdonaldtrump has such a high TF-IDF that it manages to make it to the top 20-word list. With the entry of GovMikeHuckabee, word falls out of the top 20 list again as he hasn't used that word in his tweets thereby decreasing it's score. The word "abolish" is another example which follows a similar pattern resulting in oscillating significance.

The scenarios mentioned above strengthen our stance behind the proposed methodology that every profile gets a fair chance of being clustered regardless of its distance from the user profile. This principle when extended, enables profiles which are further down the initially allocated distance list from Table 4, to get clustered out of order at a much sooner iteration. For instance, let us take the profile "GavinNewsom" as an example. According to Table 4, it is much closer to the user profile(HillaryClinton) than newtgingrich. But newtgingrich enters the cluster at iteration 3 which is way sooner than GavinNewsom's entry. Though GavinNewsom might share a lot of similar words with the user profile(HillaryClinton), newtgingrich shares more words with the existing cluster which triggers its quicker entry.

## 9 Conclusion and Future Work

In this paper, we proposed a novel way of finding the influential words between two profiles of a social media community like Twitter, which in turn, can be extended to any documents, articles on the web. We propose this idea for finding the relationship between a set of profiles using distance based or chronological order based and visualize them using *word graph*. This not only enables us to find the similar profiles but goes to the finest level and finds the words that are responsible for the similarity of profiles. This enables to trace the connection of profiles through word path (list of words). These systems can be used to classify a large number of user profiles in a social media environment. One of the limitations of the current methodology is that it cannot comprehend the same word in different forms because of the lack of understanding of semantics. In the future, we would use ontology and semantic based word recognition to prevent each word in different forms from appearing as different terms causing word redundancies.

# A Document Modeling

Example of document modeling.

TF: Term Frequency, which measures how frequently a term occurs in a document. Since every document is different in length, it is possible that a term would appear much more times in long documents than shorter ones. Thus, the term frequency is often divided by the document length (the total number of terms in the document) as a way of normalization. IDF: Inverse Document Frequency, which measures how important a term is. While computing TF, all terms are considered equally important. However, it is known that certain terms such as ”is,” ”of” and ”that,” may appear a lot of times but have little importance.

Document 1: data mining and social media mining
Document 2: social network analysis
Document 3: data mining

**Table 1.** Normalized TF of terms in document 1

| | data | mining | and | social | media |
|---|---|---|---|---|---|
| TF | 0.16 | 0.33 | 0.16 | 0.16 | 0.16 |

**Table 2.** Normalized TF of terms in document 2

| | social | network | analysis |
|---|---|---|---|
| TF | 0.16 | 0.33 | 0.16 |

**Table 3.** Normalized TF of terms in document 3

| | data | mining |
|---|---|---|
| TF | 0.5 | 0.5 |

**Table 4.** IDF of terms in corpus

| Terms | IDF |
|---|---|
| data | 1.176 |
| mining | 1.176 |
| and | 1.477 |
| social | 1.176 |
| media | 1.477 |
| network | 1.477 |
| analysis | 1.477 |

**Table 5.** Term with top IDF scores in each document

| Document | Top Words | TF-IDF |
|---|---|---|
| Document 1 | mining | 0.388 |
| Document 2 | network | 0.487 |
| Document 3 | data | 0.588 |

Table 1, Table 2 and Table 3 shows us the nomalized term frequency of each term in documents 1,2, and 3 respectively. Table 4 shows us the IDF of all terms in document 1,2 and 3. Table 5 shows us the TF-IDF of top words from document 1,2 and 3. From Table 5, it is evident that TF-IDF is high for the important words in the document and how stopwords are ignored.

## B Influential Words based on Chronological order.

. Instead of using the distance as metric for the entry of the profile into the cluster, the chronological order of entry of profiles into the cluster is taken. The chronological entry is adopted to trace the influential words a profile possess that attracted a potential follower assuming the social media adopts the recommendation of users to follow based on the influential words between profiles we proposed. The technique can be used to blow down a Twitter profile with lot of followers to list of words and understand the relationship between that profile and followers.

**Table 6.** Top 3 influential words between cluster and incoming profile(Chronological Based)

| EC | IP | Top 3 words | RC | ITR |
|---|---|---|---|---|
| HillaryClinton | GovPenceIN | govpencein<br>indiana<br>governor | I | 1 |
| I | WhoopiGoldberg | romney<br>people<br>really | II | 2 |
| II | jemelehill | laughing<br>something<br>always | III | 3 |
| LXXIV | paulwaugh | election<br>abbott<br>news | LXXV | 75 |
| CIII | taylorswift13 | lenadunham<br>theellenshow<br>mariska | CIV | 104 |
| CXXII | TwistedBacteria | science<br>stories<br>disease | CXXIII | 123 |
| CXLIII | OwenJones84 | racism<br>defeat<br>leadership | CXLIV | 144 |
| CCLXVII | KingJames | kingjames<br>brother<br>favorite | CCLXVIII | 268 |

This metric if used properly, would enable us to decompose a complex profile with a large follower-base like that of celebrities and detect the top influential words. By doing so, we can perform a detailed analysis on why people follow celebrities and which are the keywords that make a difference. Public relation officers and campaign managers for political candidates can use this analysis to target voting blocks.